\documentclass[12pt,preprint]{aastex}
\newcommand{\bb}{\begin{equation}}
\newcommand{\ee}{\end{equation}}

\shorttitle{Sunspot Seismology and Spectropolarimetry}
\shortauthors{Rajaguru et al.}

\begin{document}

\title{Local helioseismic and spectroscopic analyses of interactions between acoustic waves and a sunspot} 
\author{S.P. Rajaguru\altaffilmark{1}, R. Wachter\altaffilmark{2}, K. Sankarasubramanian\altaffilmark{3}, and S. Couvidat\altaffilmark{2}} 
\altaffiltext{1}{Indian Institute of Astrophysics, Bangalore, India}
\altaffiltext{2}{W.W. Hansen Experimental Physics Laboratory, Stanford University, Stanford CA 94305, USA}
\altaffiltext{3}{Space Science Division, Space Astronomy Group, ISRO Satellite Centre, Bangalore, India} 

\begin{abstract} 
Using a high cadence imaging spectropolarimetric observation of a sunspot and its surroundings in magnetically
sensitive (Fe {\sc i} 6173 \AA~) and insensitive (Fe {\sc i} 7090 \AA~) upper photospheric absorption lines, we map the 
instantaneous wave phases and helioseismic travel times as a function of observation height 
and inclination of magnetic field to the vertical. 
We confirm the magnetic inclination angle dependent transmission of incident acoustic waves into 
upward propagating waves, and derive (1) proof that helioseismic travel times 
receive direction dependent contributions from such waves and hence cause errors in conventional flow inferences,
(2) evidences for acoustic wave sources beneath the umbral photosphere, and (3) significant differences in
travel times measured from the chosen magnetically sensitive and insensitive spectral lines.
\end{abstract}

\section{Introduction}
\label{sec:intro}

Accounting for directly observable photospheric wave evolution within a sunspot
while inferring subsurface conditions is an important and challenging problem in
the seismology of sunspots \citep{bogdanetal98}. 
The characteristic association between non-vertical magnetic fields 
and acoustic wave propagation at frequencies well below the photospheric acoustic cutoff of $\approx$ 5.2 mHz,
observed in sunspot penumbrae \citep{mcintoshjefferies06,rajaguruetal07} as well as in disparate
wave dynamical phenomena in the solar atmosphere \citep{depontieuetal04,depontieuetal05,jefferiesetal06,dewijnetal09}
has further highlighted this problem.
Several indirect influences, due to physical \citep{woodard97,bogdanetal98} as well as analysis specific 
reasons \citep{rajaguruetal06,parchevskyetal07}, of {\em p}-mode absorption \citep{braunetal87} per se 
have been shown to manifest as apparent flow and wave speed signals in local helioseismic measurements; however,
there have been none studying the direct contributions of propagating waves. 
The "inclined magnetic field effect" in helioseismic signatures \citep{schunkeretal05,zhaoetal06}
has indeed been shown to arise from magnetic-field-aligned wave motion caused by an incident acoustic wave,
but the interpretations relied on a viewing-angle-dependent geometric relation between the magnetic field and
wave motion, whose phase shifts are independent of those arising from wave progression in height. 
Viewing-angle-dependent changes in observation height too could cause different
wave phases at different positions within a sunspot and potentially could be misinterpreted
as the above effect, as suggested by \citet{rajaguruetal06}.

In this Letter, we derive explicit observational proofs for helioseismic contributions from propagating 
waves within a sunspot and from acoustic sources located beneath its umbral photosphere.
We also show significant differences in travel times measured using velocity data from 
magnetic and non-magnetic lines.

\section{Observations and Analysis Methods}
We performed imaging spectropolarimetry using the Interferometric BI-dimensional Spectrometer (IBIS) 
installed at the Dunn Solar Telescope of the National Solar Observatory, Sac Peak, New Mexico, USA.  IBIS has spectral and 
spatial resolutions of 25 m\AA~ and 0".165, respectively, and has a 80" diameter ($\approx$60 Mm) circular field of view (FOV). 
We observed a medium sized sunspot (NOAA AR10960, diameter 
$\approx$ 18 Mm) located close to the disk center (S07W17) on 2007 June 8. Our observations involved
scanning and imaging in all the Stokes profiles ($I,Q,U,V$) of magnetic Fe {\sc i} 6173.34 \AA~ and in Stokes $I$ of non-magnetic
Fe {\sc i} 7090.4 \AA~, with a cadence of 47.5 s. A 7 hr continuous observation was chosen for our analysis. 
The spectral images were dark subtracted, flat-fielded, and re-registered to remove atmospheric distortions, which were 
derived from white-light images recorded simultaneously. 
The spectral calibration consisted of removing a quadratic center-to-edge wavelength dependence, and correcting for the 
transmission profiles of the prefilters. Polarization calibration of the magnetic line data was based on the telescope 
and instrument polarization matrices \citep{cavallini06}.


Similar to \citet{rajaguruetal07}, we extract line-of-sight (LOS) velocities of plasma motions 
within the line forming layers from the Doppler shifts of line bisectors. 
We use 10 bisector levels with equal spacing in line intensity, ordered from the line core (level 0)
to the wings (level 9), and derive 10 velocity data cubes, $v_{i}(x,y,t)(i=0,...,9)$, for each line.
For the magnetic line, we use the average of bisector
velocities from the left ($I+V$) and right ($I-V$) circular polarization (CP) profiles 
\citep{sankar-rimmele02,dtiniesta03} and those from the $I$ profile for the non-magnetic line.
The 10 bisector levels span the height range within the line formation region in an unique
one-to-one way. 
Based on the Maltby-M umbral model atmosphere \citep{maltbyetal86}, 
the Fe {\sc i} 6173.34 \AA~ line formation is reported to span a height range of
20 km (wings) to 270 km (line core) above continuum optical depth $\tau_{c}$=1 level
\citep{nortonetal06}, and a very similar range for Fe {\sc i} 7090.4 \AA~ \citep{strausetal08}.
We choose about 1 hr long observation from the best seeing interval (the first 3 hr),
and do Milne-Eddington (M-E) inversions of the (temporal) average of Stokes profiles of the magnetic line 
to obtain magnetic field $B(x,y)$, its LOS inclination $\gamma(x,y)$ and 
azimuth $\psi(x,y)$ \citep{sku-lites87}. Because of low polarization signals, the above inverted quantities
are noisy outside of the sunspot (see Figure 1), and we cut out only the spot region for use in our
analysis. 

\subsection{Instantaneous Wave Phases and Helioseismic Travel Times}

Instantaneous wave phases in the form of phase shifts
$\delta\phi_{i,0}(\nu)$=Phase[${\mathbf V_{i}}(\nu){\mathbf V^{*}_{0}}(\nu)$], where $\nu$ is the cyclic frequency of a wave and
${\mathbf V}$ is the Fourier transform of $v$, due to wave progression between two heights 
corresponding to any one of the bisector levels $i=1,2,...,9$ and level $0$ (the top most layer) are
calculated \citep{rajaguruetal07}. Since we want to study mainly the $p$-modes and compare $\delta\phi$ with helioseismic
travel times, we filter out the $f$-mode.
We take median values of $\delta\phi$ over the $p$-mode band (2 - 5 mHz) or over bands of 1 mHz full width at half-maximum (FWHM)
centered at every 0.25 mHz (to study any frequency dependence). 
Signals over space are studied using $\gamma$ or LOS magnetic field $B_{LOS}$. For this
work we focus on studying the $\gamma$ dependence of $\delta\phi$ and use 
3$^{\circ}$ bins in $\gamma$ (see Section $\ref{subsect:errors}$ below). 

The 10 different data cubes from each line are run through a standard $p$-mode time-distance analysis procedure
in center-annulus geometry \citep{rajaguruetal04}. In addition, we apply frequency filters the same way as for $\phi$.
Travel time maps are calculated for five travel
distances $\Delta = 6.2, 8.7, 11.6, 16.95,$ and $24$ Mm (see \citet{couvidatbirch09} for details on phase speed
filters). For this Letter, we focus on analyzing the results for $\Delta = 16.95$ Mm, because, given the sizes 
of observed region (radius $\approx$ 29 Mm) and the spot (radius $\approx$ 9 Mm), this is the optimum $\Delta$ that
facilitates distinguishing clearly the ingoing and outgoing waves in the sense of their interactions with the spot. 
We also perform a double-skip annulus-annulus
geometry measurement, for $\Delta = 16.95$ Mm, which avoids use of oscillation signals within the sunspot, for the
diagnostic checks presented in Section 4. 
Height dependent contributions to outgoing and ingoing
phase travel times $\tau^{+}$ and $\tau^{-}$ from within the line forming layers are determined using 
$\delta\tau^{\pm}_{i,0}=\tau^{\pm}_{0}-\tau^{\pm}_{i}$ ($i=1,...,9$).
To facilitate comparisons with $\delta\phi$, we average $\delta\tau^{\pm}_{i,0}$ too over 3$^{\circ}$ bins of $\gamma$.
All pixels outside of 10 Mm radius centered on the spot, for convenience (in Figures 2 - 4),
are assigned a $\gamma$ value of 90$^{\circ}$; all wave quantities averaged over these set of pixels are taken
as that of 'quiet-Sun', and a subscript $q$ is used, where necessary, to identify them explicitly.

\subsection{Error Analysis}
\label{subsect:errors}
The oscillation signals are inherently stochastic due to such nature of acoustic sources and that of the background medium.
A typical measurement of a wave quantity, hence, here either in $\delta\phi$ or $\delta\tau$, carries a random error. 
Assuming that all pixels with the same $\gamma$ or within a small range of $\gamma$ comprise independent measurements
of the same $\delta\phi$ and $\delta\tau$, we take the mean over these pixels as our best estimate and study its variation
against $\gamma$. 
Error estimates for $\gamma$ from M-E inversions fall in the range of 0$^{\circ}$.85 - 2$^{\circ}$, with mean values of
1.5$^{\circ}$ over the umbra and 1$^{\circ}$ over the penumbra.
A bin size of 3$^{\circ}$ in $\gamma$, referred to in previous subsection, is found to be optimal
to accumulate a statistical sample of measurements while being small enough to not bias them through their variation against
$\gamma$ itself. So, error estimates
for the means $\delta\phi$ or $\delta\tau$ are their standard errors given by $\sigma /\sqrt{n}$, where $\sigma$ is the 
standard deviation in $n$ number of measurements (i.e. pixels falling within a given bin in $\gamma$).

\section{Origin and Seismic Contributions of Propagating Waves}

Oscillation signals in photospheric Doppler velocities, in general, would consist of evanescent waves caused by
the $p$-modes trapped below it and various propagating waves traveling in different directions. 
The $\delta\phi_{i,0}$ arise only due to the latter propagating waves.
The surface- ($f$-mode) and atmospheric-gravity waves also show vertical phase propagation
in the photospheric layers, but we have filtered them out here in our analysis.
In general, $\delta\phi_{i,0}$ should receive contributions from waves locally generated and those
generated elsewhere (e.g., from the quiet-Sun) but get "converted" by the magnetic field to propagate 
upwards upon incident on it. On the other hand, the ingoing travel times $\delta\tau^{-}_{i,0}$ get contributions
solely from the latter helioseismic ones, which in our current analysis case are those traveling from
a distance of $\Delta=$ 16.95 Mm in regions surrounding the spot. 
We show in Figure 2 $\delta\phi_{8,0}$ and $\delta\tau^{-}_{8,0}$, due to wave evolution within the 
region bounded by the wing (level 8) and core (level 0) formation heights, against $\gamma$.
The $\nu$ values marked in the panels of Figure 2 are the central
frequencies of 1 mHz band filters used. Keeping in mind that $\delta\phi_{8,0}$ have contributions from a larger
set of waves (as discussed above), results in Figure 2(a) for the magnetic line show a surprising 
amount of correlation between the two measurements, and moreover exhibit a strikingly similar $\gamma$ dependence.
These results immediately reveal several interesting aspects of magnetic field - acoustic wave interactions:
(1) first of all they confirm that helioseismic waves incident on the sunspot see themselves through to higher
layers of its atmosphere with a striking dependence on $\gamma$: a coherent transmission
of incident waves happen, peaking around $\gamma \approx$ 30$^{\circ}$, maintaining
a smooth evolution of time-distance correlations; (2) remembering that CP profiles of the magnetic line have maximum
sensitivities for velocities within vertical magnetic field, it is seen that a large fraction of waves propagating upward
within such field are due to helioseismic waves originating at distant locations;
and, (3) provide direct evidences that ingoing wave travel times would cause observing height 
dependent signals in flow inferences from travel time differences. 
The non-magnetic line (Figure 2(b)) yields
very little correlations between $\delta\phi_{8,0}$ and $\delta\tau^{-}_{8,0}$; however, the helioseismic measurements 
$\delta\tau^{-}_{8,0}$ agree well with that in Figure2(a), except at high $\nu$, thus reinforcing inferences (1) and (3).
We speculate that there are substantial wave motions, locally generated, perhaps within the non-magnetic gaps or weakly
magnetized penumbral region, whose signatures are missed in the CP profiles of the magnetic line.
At $\nu \geq$ 4 mHz, helioseismic signatures within the sunspot get markedly different in magnetic and non-magnetic lines.

To affirm the reader that the signals analyzed in Figure 2 (as well as other Figures) are due to the height evolution
of wave phases and not due to any other wave correlations, random or spurious, in Figure 5 we show 
maps of $\delta\phi_{i,0}$ and $\delta\tau^{-}_{i,0}$ for the full $p$-mode band (2 - 5 mHz), whose variations against 
$i=1,2,...,8$ are in the $x$-direction of the maps. The gradual increase in the values of $\delta\phi_{i,0}$ and $\delta\tau^{-}_{i,0}$
with that in height separation is obvious in this figure. We further note that halving the bin size
in $\gamma$ to 1.5$^{\circ}$ changes negligibly the mean values $\delta\phi$ and $\delta\tau$ and their variation
against $\gamma$ studied in this work, but increases their error estimates by roughly $\sqrt{2}$ times.

\section{Wave Sources Beneath Umbral Photosphere}
Outgoing waves at a given measurement location, in general, would consist of those locally generated 
and those generated elsewhere undergoing reflection at the photosphere directly below it. These latter component
would be seen in neither $\delta\phi_{i,0}$ nor $\delta\tau^{\pm}_{i,0}$, as they are evanescent at the observing height.
For locally generated waves, circular wavefronts from a source, while their upward propagating parts see 
themselves up through the magnetic field, would cause outgoing wave correlations yielding distinct 
signatures in $\delta\tau^{+}_{i,0}$ (see Figure 3(b)). Results in Figure 3(a), for $\delta\tau^{+}_{8,0}$ from
both the magnetic and non-magnetic lines, do indeed provide such a
diagnostic: outgoing waves starting at higher height (line core) within the sunspot atmosphere 
and reaching the quiet-Sun at the chosen $\Delta$ have shorter travel times than those starting at a lower height 
(line wings) and reaching
the same quiet-Sun location; since this is simply not possible, the only explanation for this observation
is the one contained in our previous sentence and illustrated in Figure 3(b), 
viz., outgoing wave time-distance correlations are predominantly 
due to waves directly from sources just beneath the sunspot photosphere when oscillations observed within it are used. 
We also note the different $\gamma$ dependence of $\delta\tau^{+}$, as compared to that of $\delta\tau^{-}$, possibly
due to circular wavefronts from local sources having a range of incident angles with the magnetic field. 

To confirm the above and to check the extent of contributions from magnetic patches surrounding the spot, we 
measured travel times in double-skip annulus-annulus geometry that avoids waves observed within the sunspot. 
Here, diametrically opposite points on the annulus are correlated and azimuthally averaged; hence only the
mean travel times are measured and assigned to the center points, directly beneath which the waves 
reflect between their two skips.
The results in Figure 4 show that $\delta\tau^{ds}_{8,0}$ are indeed small, compared to $\delta\tau^{+}_{8,0}$. Hence,
the small scale magnetic patches surrounding the spot do not contribute much to $\delta\tau^{ds}_{8,0}$ and hence to
$\delta\tau^{\pm}_{i,0}$ too, validating
our inferences above and in Section 3. As to magnitudes of $\delta\tau^{+}_{8,0}$ (Figures 3a and 4), it is interesting
to note that they are about thrice those of $\delta\tau^{-}_{8,0}$ (Figure 2) (in the main $p$-mode band of 2 - 4 mHz).
This difference could arise from non-circular expansion of wavefronts possibly due to two causes: (1) the differences 
in physical conditions, due to material flows as well as sound speed, in the wave path regions H$_{c}$-H$_{w}$ and 
H$^{\prime}_{c}$-H$^{\prime}_{w}$ (refer to Figure 3(b)), 
and (2) the motion of the sources beneath the umbral photosphere. In the lower panel of Figure 4, we compare
mean travel time perturbations measured from line core and wing bisector velocities, $\tau^{m}_{0}$-$\tau^{m}_{q0}$ and
$\tau^{m}_{8}$-$\tau^{m}_{q8}$, with those of half the double-skip travel times, ($\tau^{ds}_{8}$-$\tau^{ds}_{q8}$)/2, 
from wing bisector velocities. Interestingly, almost all the differences between double- and single-skip travel times
appear to come from the line formation layers, and hence are observation height dependent. These results
also show that height difference between the source location and the wave reflection layer beneath the spot are very small.
 
\section{Discussions and conclusion}

Almost all time-distance helioseismic analyses proceed under the working assumption that wave signals
at observation heights are evanescent and hence oppositely directed wave paths
involving photospheric reflections at two separated points are of identical path length. This assumption is basic to the
inferences on flows and wave speed from travel time differences and mean, respectively.
In an early theoretical study, accompanied by attempts to model the helioseismic observations 
of \citet{braun97}, \citet{bogdanetal98} showed the influences of both the $p$-mode forcing of, and spontaneous emissions by, 
sunspots on acoustic wave travel times.
Our analyses here have yielded transparent observational proofs for both effects, for the first time,
with important new perspectives: (1) the process of transformation of incident acoustic waves into propagating
(magneto)-acoustic waves up through the magnetic field happen in a coherent manner allowing a smooth evolution
of time-distance correlations and, in agreement with several recent theoretical and numerical studies \citep{cally05,crouchandcally05,
schunkeretal06}, this process depends on the inclination angle ($\gamma$) of magnetic field to the vertical,
and (2) outgoing waves from acoustic sources located just beneath the
sunspot photosphere add important additional contributions for both mean travel times and differences.
Our results have also shown observational prospects for
consistently accounting for the above effects in sunspot seismology, viz. the indispensability of imaging spectroscopy to
extract wave fields so as to be able to correctly account for the wave evolution within the directly
observable layers of sunspot atmosphere. Current limitations in making such observations over large enough FOV 
do not allow us to perform seismic inversions reliably. However, the analysis methods followed here 
point ways to a consistent and much improved observational determinations of structure and flows beneath sunspots
once our instrumental capabilities improve. These observational avenues also promise a close scrutiny of
various theoretical ideas and models of acoustic wave - magnetic field interactions and those of the associated
MHD waves and their propagation characteristics. 
 
\acknowledgments
R. Wachter and S. Couvidat are supported by NASA Grant NAS5-02139 to HMI/$SDO$ project 
at Stanford University. We thank K. Reardon and A. Tritschler for help in observations and data calibration.
We thank Baba Varghese (IIA, Bangalore) for help in drawing Figure 3(b).

\newpage

\begin{figure}
\epsscale{1.1}
\centering
\figurenum{1}
\plotone{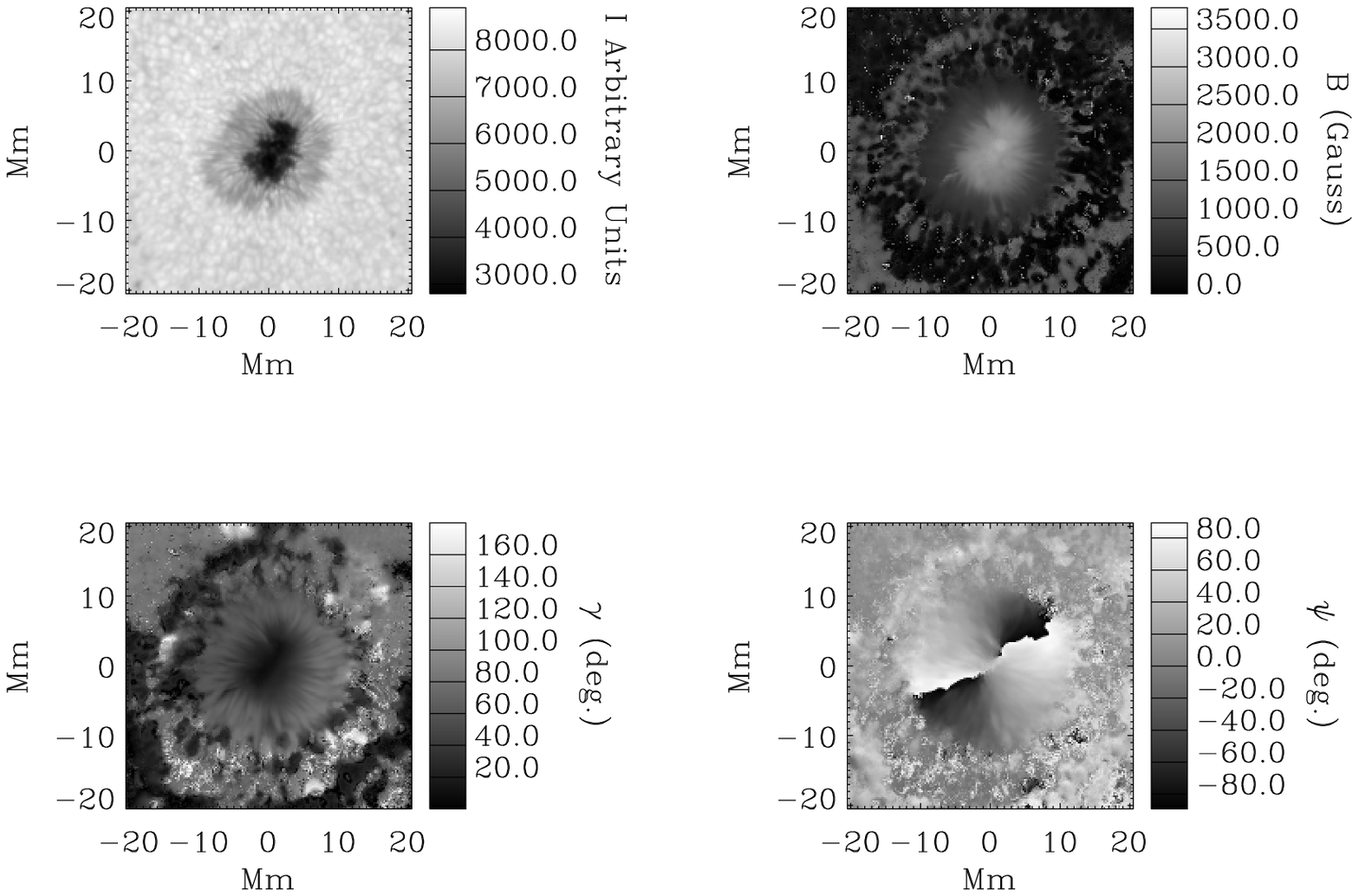}
\caption{Maps of continuum intensity $I_{c}$, and magnetic field $B$, its azimuth $\psi$, and LOS
inclination $\gamma$ over the sunspot region from M-E inversions of Fe {\sc i} 6173 \AA~ Stokes profiles.}
\label{fig:1}
\end{figure}

\begin{figure*}
\centering
\figurenum{2}
\plottwo{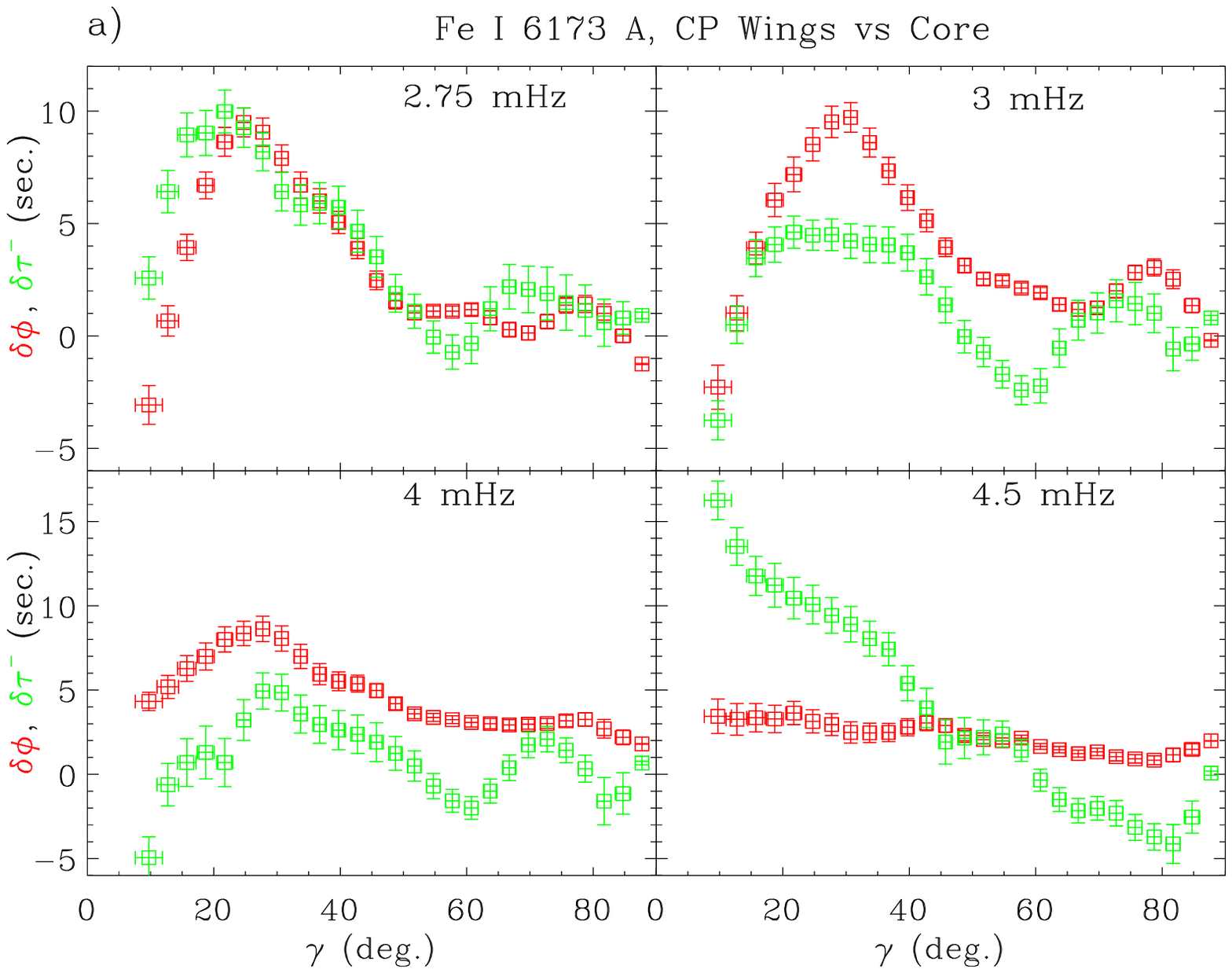}{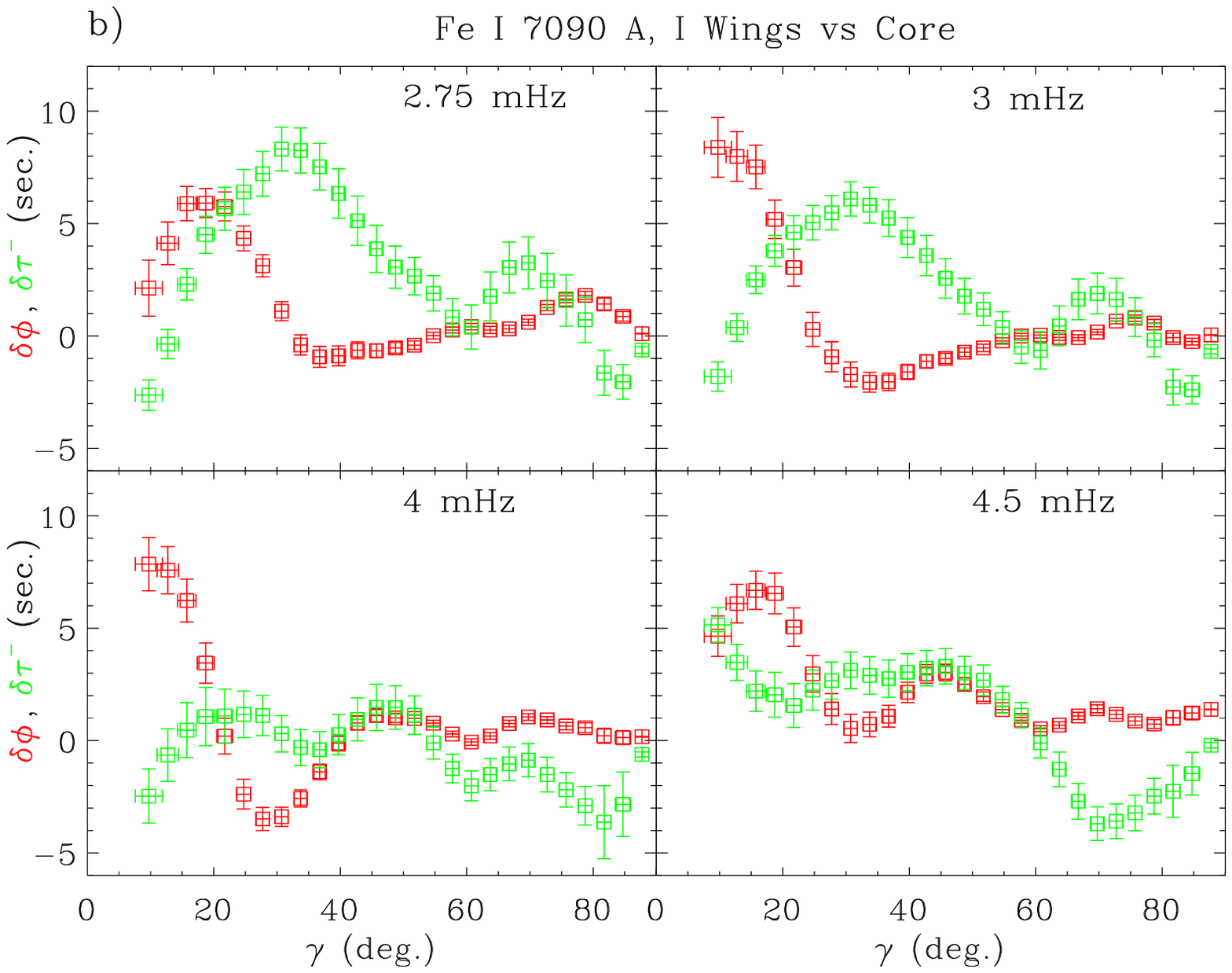}
\caption{Instantaneous phase shifts, $\delta\phi_{8,0}(\nu)$ (in red), and changes in ingoing wave travel times,
$\delta\tau^{-}_{8,0}$ (in green), due to wave propagation between the formation heights of wings (20 km) and core (270 km)
of Fe {\sc i} 6173 \AA~({\em panel a}), and of Fe {\sc i} 7090 \AA~({\em panel b}) against $\gamma$ of B.}
\label{fig:2}
\end{figure*}
\begin{figure}[ht]
\epsscale{1.1}
\centering
\figurenum{3}
\plottwo{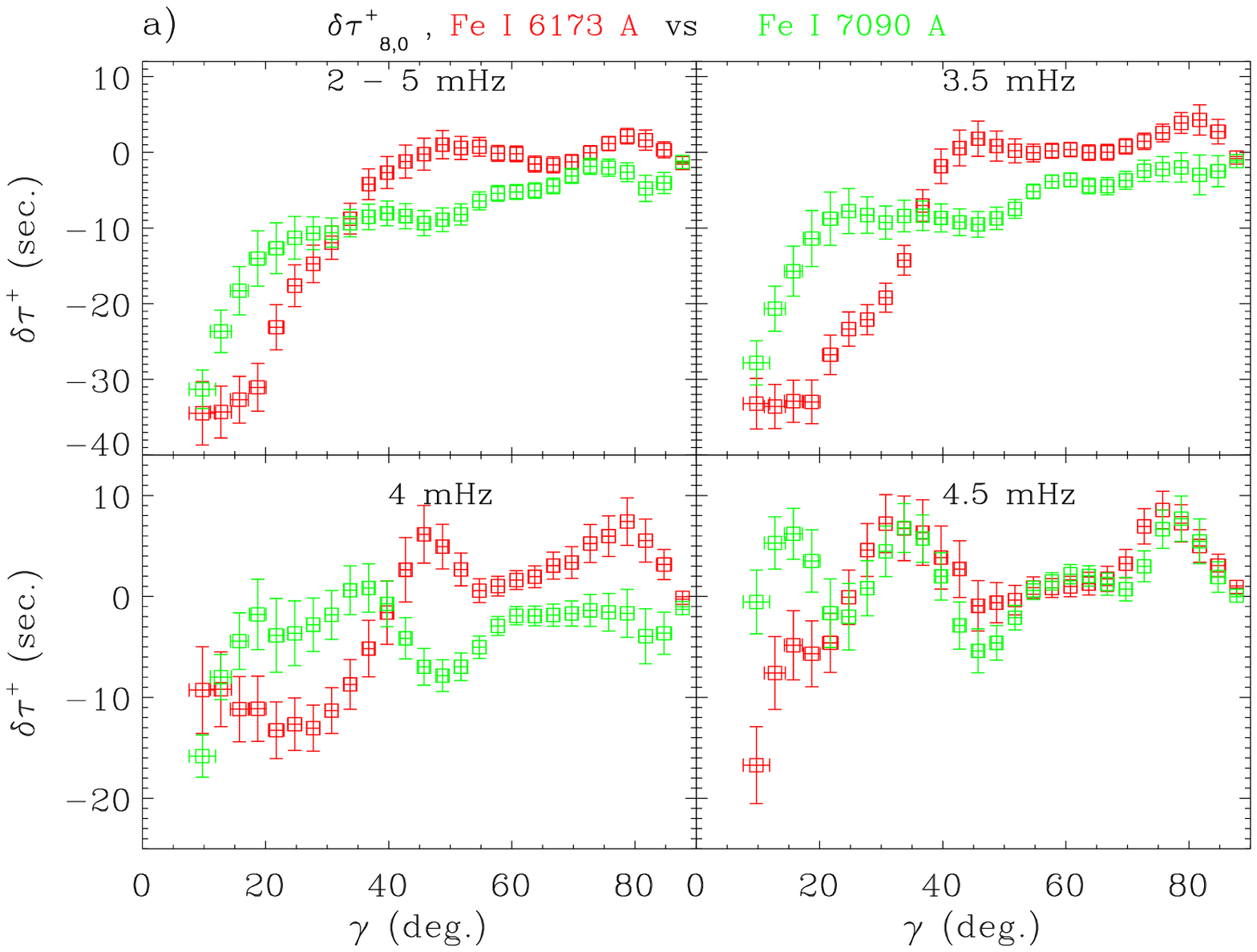}{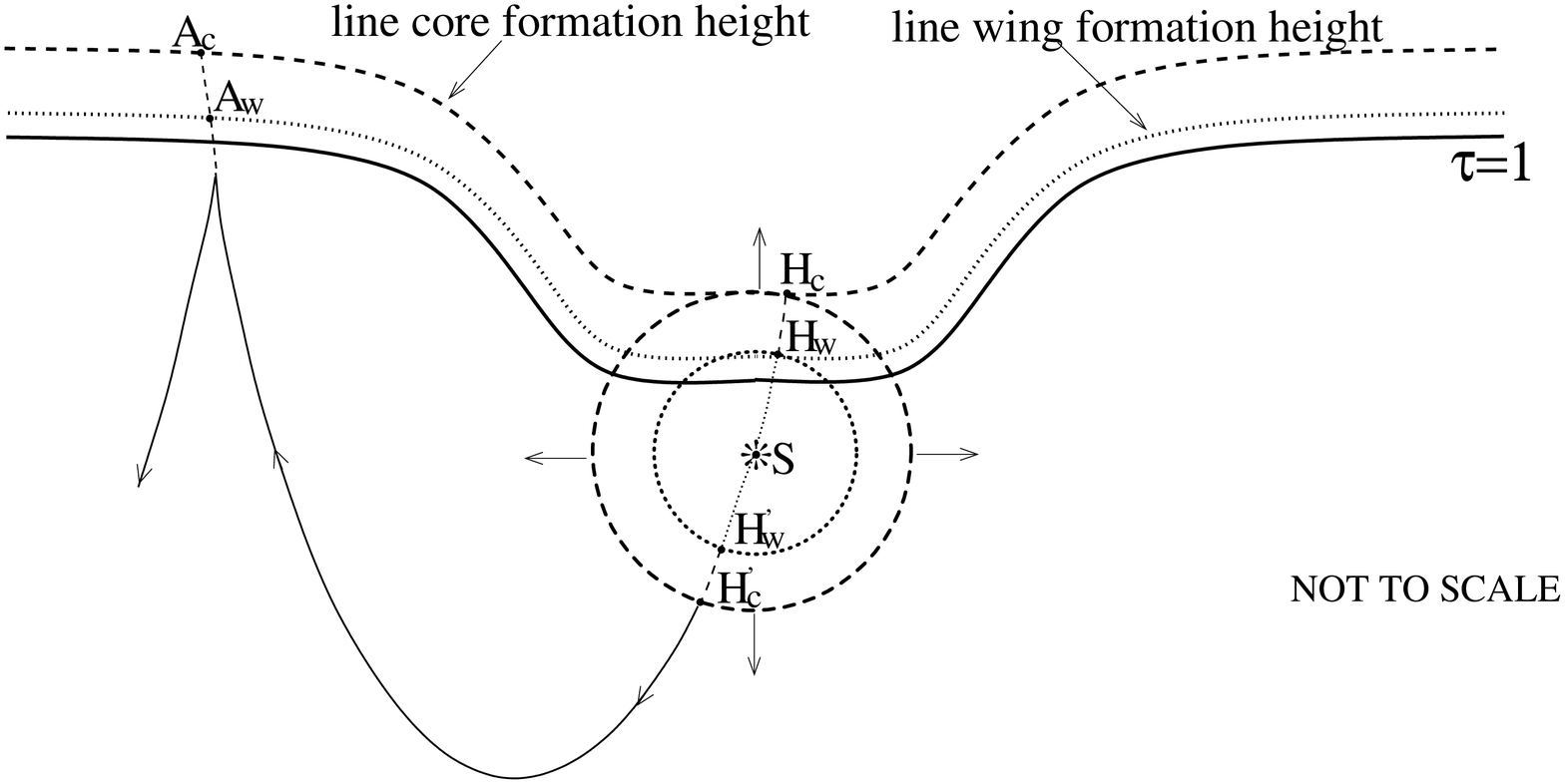}
\caption{({\em Panel a}) Changes in outgoing wave travel times, $\delta\tau^{+}_{8,0}$, due to wave propagation between the
formation heights of the wings (20 km) and the core (270 km) of Fe {\sc i} 6173 \AA~ (in red) and 7090 \AA~ (in green)
lines as a function of $\gamma$ of B. ({\em Panel b}): an illustration depicting wavefronts from acoustic sources beneath the
umbra, wave paths and line formation heights (see text for details).}
\label{fig:3}
\end{figure}


\begin{figure}[ht]
\epsscale{1.1}
\centering
\figurenum{4}
\plotone{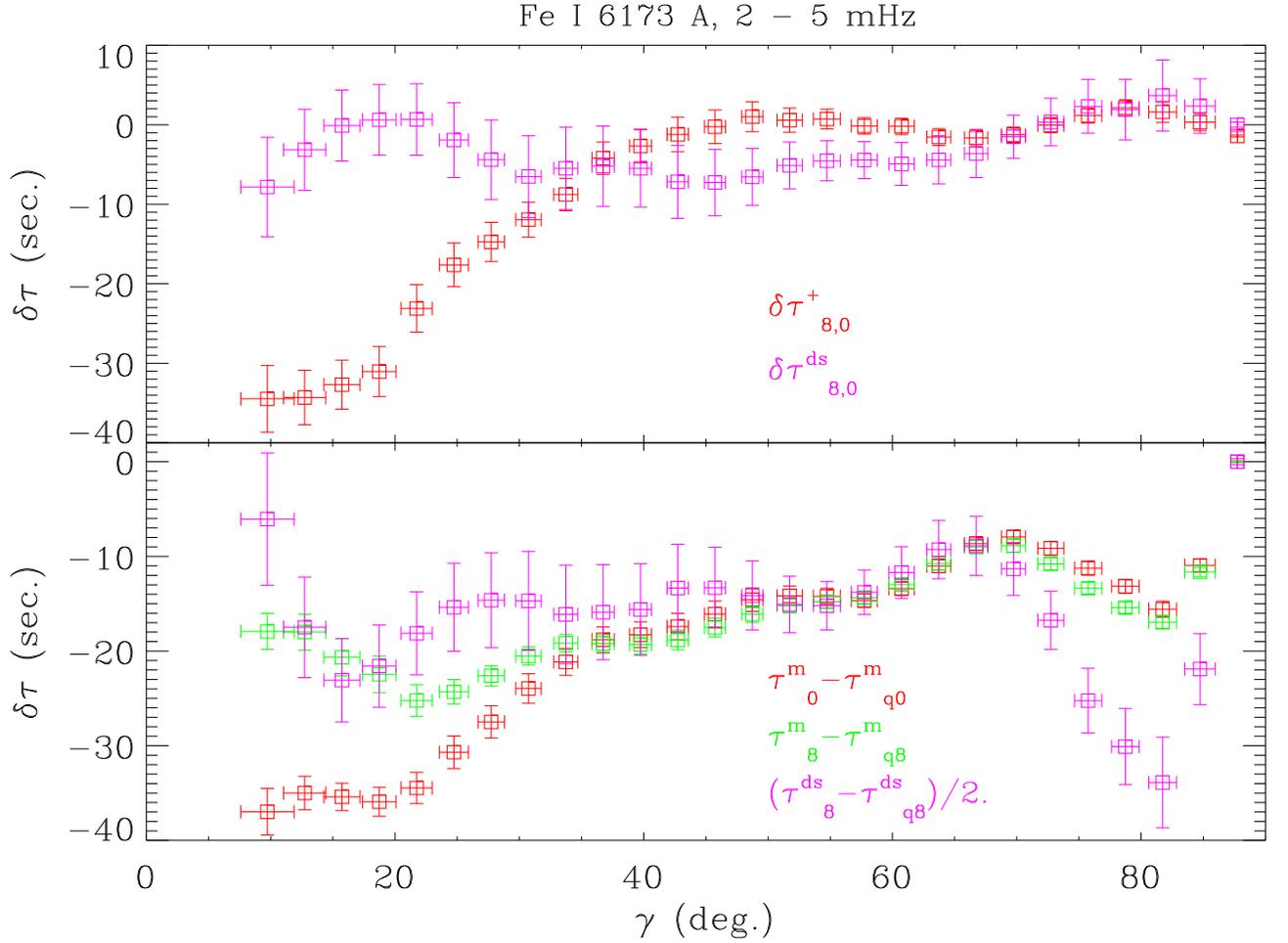}
\caption{{\em Upper panel}: comparing outgoing wave travel times $\delta\tau^{+}_{8,0}$ and double-skip mean
travel times $\delta\tau^{ds}_{8,0}$ from within the line formation region. {\em Lower panel}:
changes, with respect to quiet-Sun, in half of double-skip travel times (violet)
and in mean travel times at formation heights of line wings (green) and core (red).}
\label{fig:4}
\end{figure}

\begin{figure}[ht]
\epsscale{0.9}
\centering
\figurenum{5}
\plotone{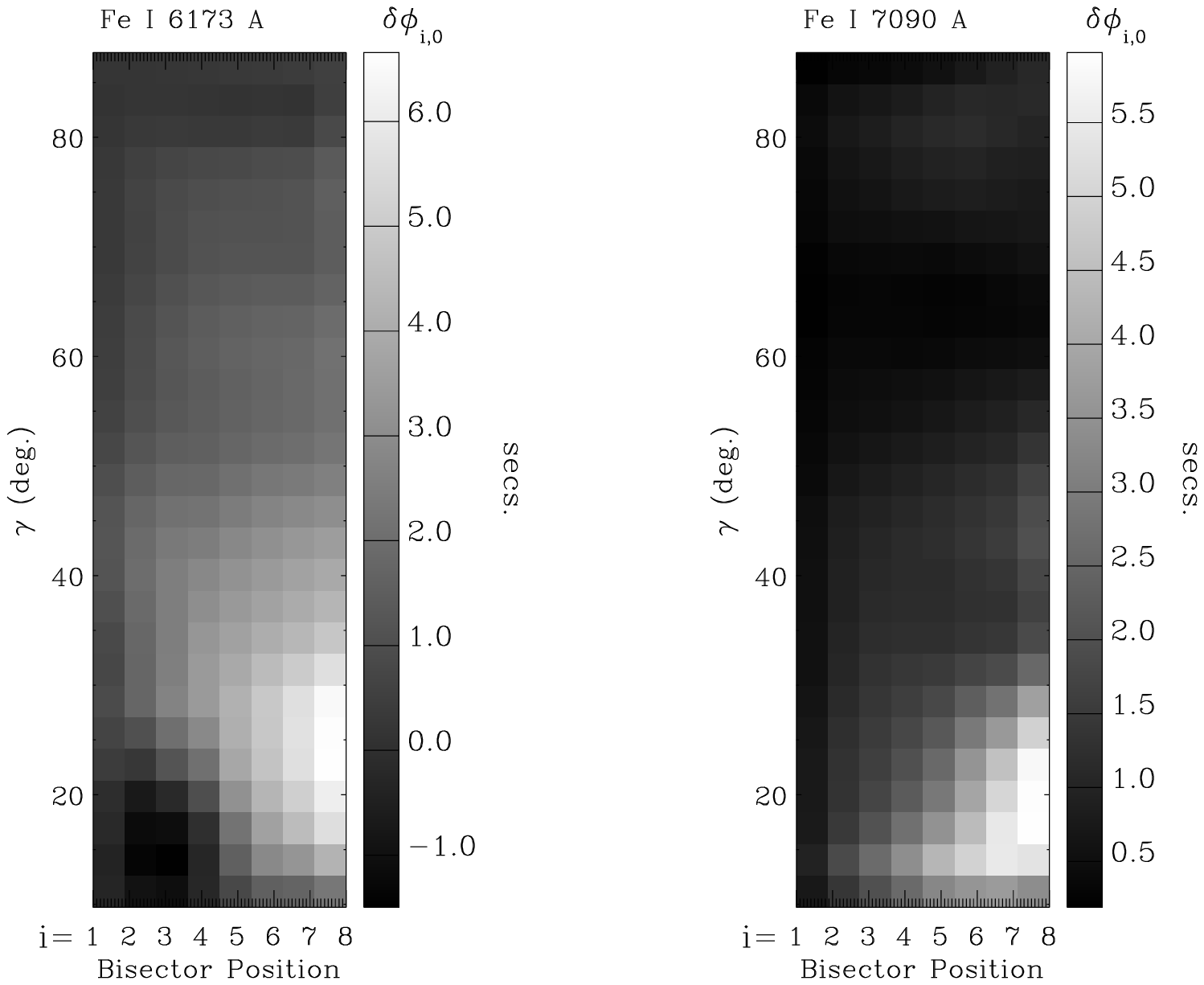}
\plotone{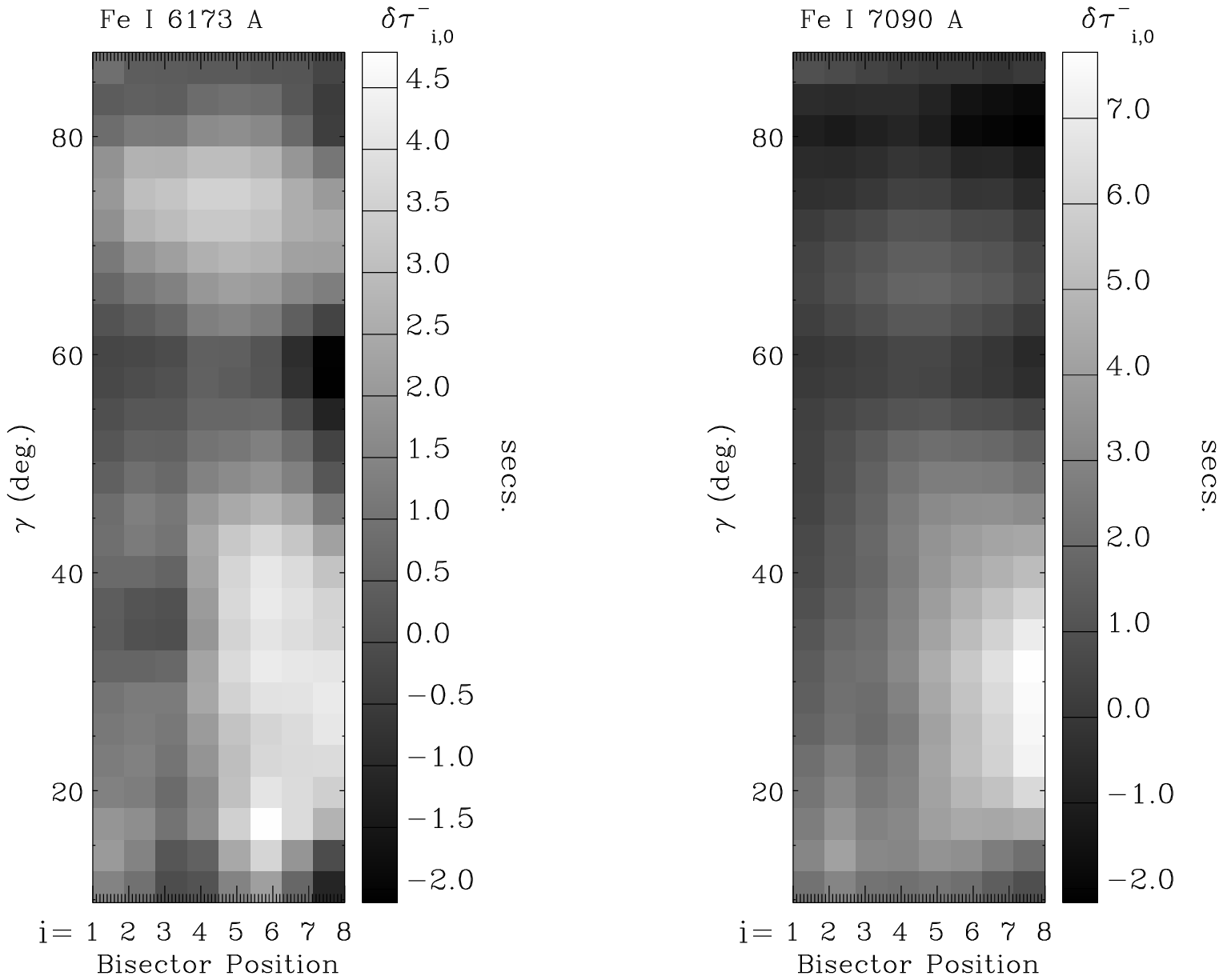}
\caption{Maps of $\delta\phi_{i,0}$ ({\em top row}) and $\delta\tau^{-}_{i,0}$ ({\em bottom row})  for
the full $p$-mode band (2 - 5 mHz). The variation in the x-direction of the above maps is for the
bisector levels $i=1,2,...,8$ corresponding to progressively deeper locations in the sunspot
atmosphere with respect to $i=0$, which is for the line core forming at 270 km above the photosphere.}
\label{fig:5}
\end{figure}

\end{document}